\documentclass[
    ,final            % use final for the camera ready runs
  ]
  {aipproc}

\layoutstyle{6x9}

\def\lsim{\raise0.3ex\hbox{$<$\kern-0.75em\raise-1.1ex\hbox{$\sim$}}}
\def\gsim{\raise0.3ex\hbox{$>$\kern-0.75em\raise-1.1ex\hbox{$\sim$}}}

\begin{document}

\title{QCD Thermodynamics with an almost realistic quark mass
  spectrum\\[-3cm] \hspace*{11.9cm}{\normalsize \tt BNL-NT-06/6}\vspace*{2.4cm}}


\author{C. Schmidt for the RBC-Bielefeld Collaboration}{
  address={Brookhaven National Laboratory, Physics Department, Upton, NY, 11973, USA}
}

\begin{abstract}
We will report on the status of a new large scale calculation of thermodynamic
quantities in QCD with light up and down quarks corresponding
to an almost physical light quark mass value and a heavier strange quark mass.
These calculations are currently being performed on the QCDOC
Teraflops computers at BNL. We will present new lattice calculations of the
transition temperature and various susceptibilities reflecting properties of the
chiral transition. All these quantities are of immediate interest for
heavy ion phenomenology.
\end{abstract}

\maketitle

\section{Introduction and Lattice Setup}
The lattice action we use is designed for finite temperature Monte Carlo
Simulations. In the gauge sector we use a $(2\times1)$-Symanzik improvement
scheme which eliminates all cut-off effects of order ${\cal O}(a^2)$, where $a$ is
the lattice spacing. For the fermions we use the staggered fermion
formulation. On top of that we add an improvement term which restores the
rotational symmetry of the free quark propagator on the lattice up to order
${\cal O}(p^4)$ in the momentum $p$ \cite{Heller:1999xz}. In order to improve
the flavor symmetry, which is violated in the staggered fermion formulation, we
smear the gauge field which is used in the standard part of the fermion
action. The procedure involves either only three link terms (p4fat3) or up to
seven link terms \cite{Orginos:1999cr} (p4fat7). While the p4fat3 action was
used for thermodynamical calculation earlier
\cite{Karsch:2000ps,Karsch:2000kv}, the combination of the p4-term and seven
link smearing is used here for the first time in thermodynamic calculations
\cite{p4fat7_1,p4fat7_2}. 

We perform simulations with three degenerate quark flavors as well as with
2 light and one heavy quark flavor. The lattices have temporal extent
$N_t=4,6$, which corresponds at the critical temperature ($T_c$) to a lattice
spacing of $a\approx0.13$~fm and $0.22$~fm respectively. The lattice extent in spacial
direction is $N_s=8,16,32$. To determine the scale, we perform zero
temperature simulations on $16^3\times 32$ lattices. These calculations are
being performed on the QCDOC Teraflops computers at BNL.

\section{The Critical Temperature}
\begin{figure}
\begin{minipage}{.5\textwidth}
\begin{center}
\includegraphics[height=.64\textwidth, width=1.0\textwidth]{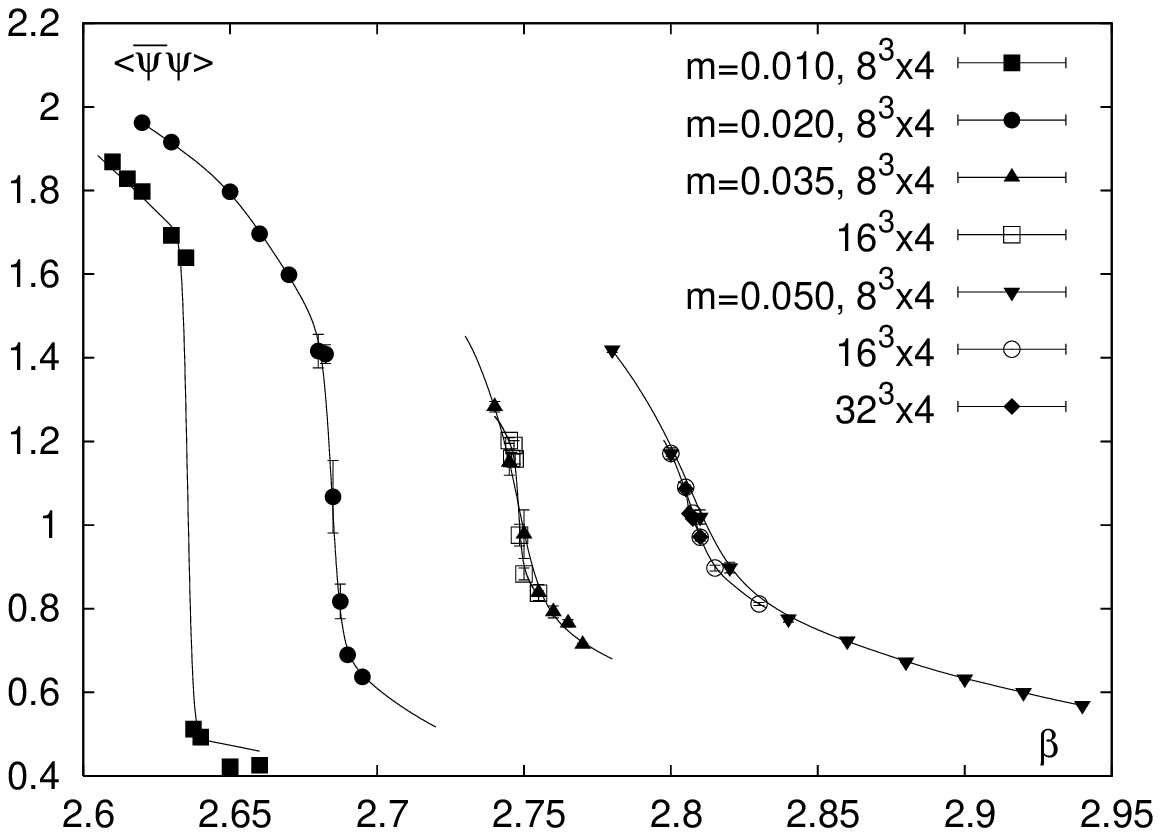}
\end{center}
\end{minipage}
\begin{minipage}{.5\textwidth}
\begin{center}
\includegraphics[height=.64\textwidth, width=1.0\textwidth]{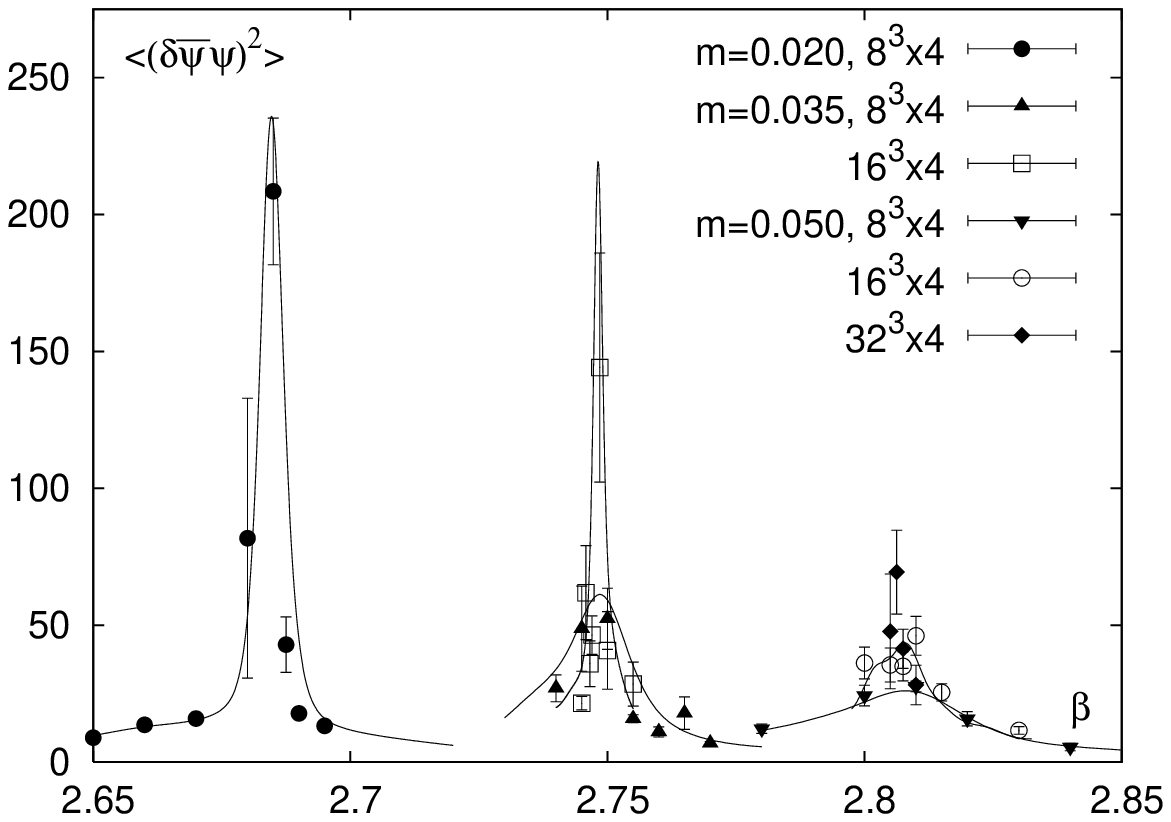}
\end{center}
\end{minipage}
\caption{The chiral condensate and (left) and the chiral susceptibility
  (right) for the p4fat7 action with three degenerate flavor and $N_t=4$. Data
  points are the individual Monte Carlo runs; solid curves are from multi-histogram
  reweighting.\label{fig_pbp}} 
\end{figure}

\begin{figure}
\vspace*{-5mm}
\begin{minipage}{.5\textwidth}
\begin{center}
\includegraphics[height=.64\textwidth, width=1.0\textwidth]{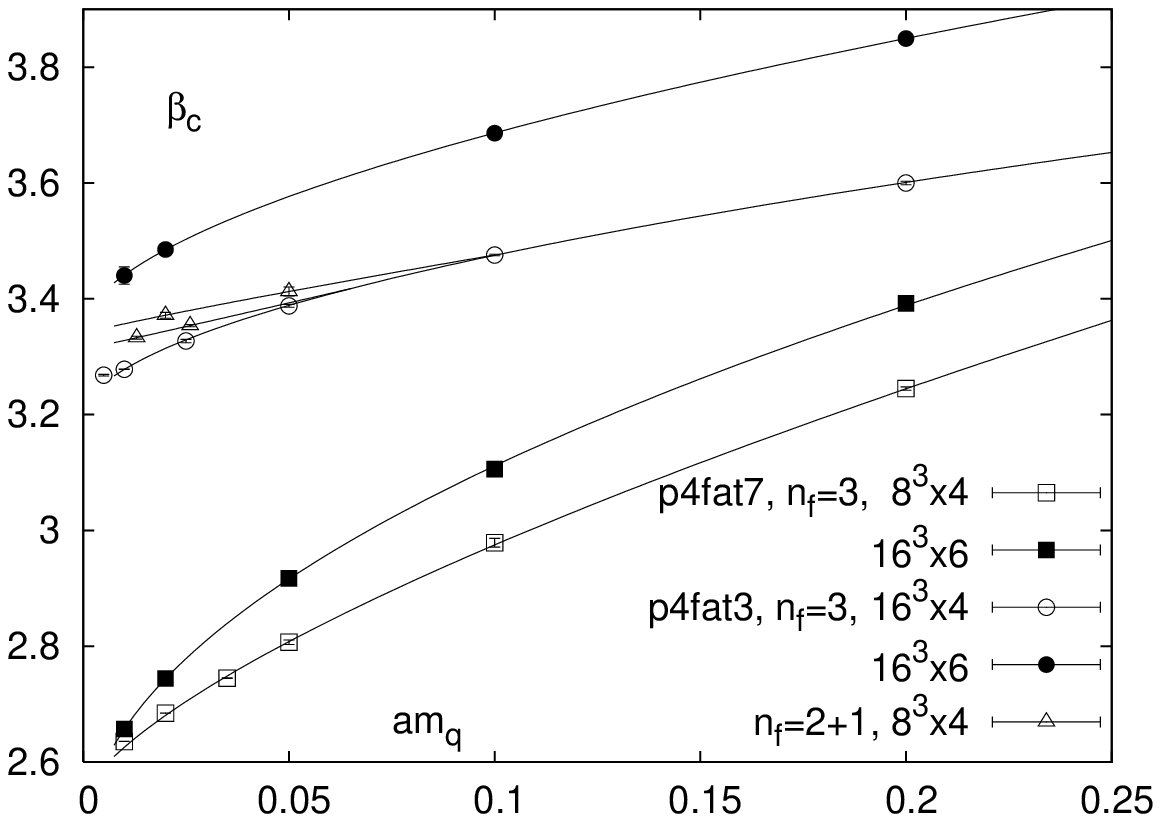}
\end{center}
\end{minipage}
\begin{minipage}{.5\textwidth}
\begin{center}
\includegraphics[height=.64\textwidth, width=1.0\textwidth]{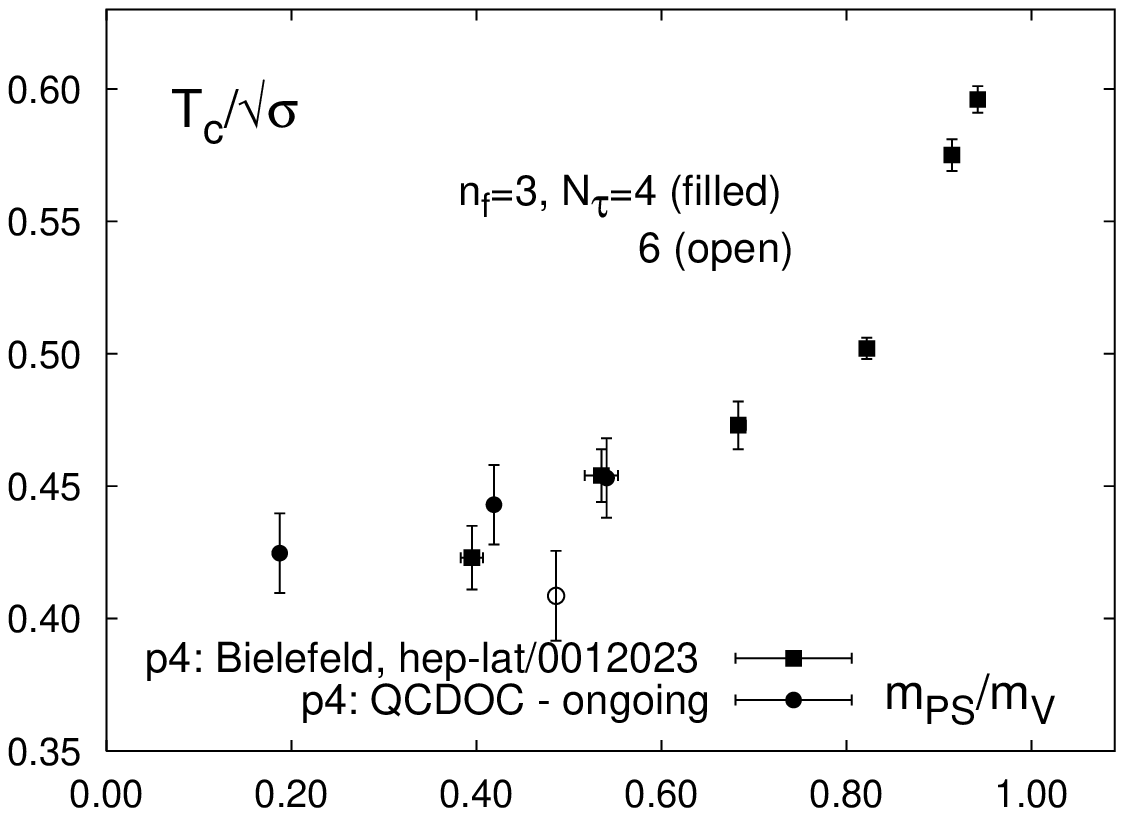}
\end{center}
\end{minipage}
\caption{The critical couplings vs. the bare quark masses for different actions
  and lattice sizes (left) and the critical temperature in units of the string
  tension ($\sigma$) as a function of the ratio of pseudo-scalar and vector meson 
  masses (right).\label{fig_betac}}
\end{figure}

The observables we calculate at present are the gauge action, the Polyakov loop and the
chiral condensate. For each of these quantities we calculate also its
susceptibility. Using the multi-histogram reweighting technique
\cite{Ferrenberg:1988yz} we combine the statistics of Monte Carlo runs at
different couplings $\beta$. Results for the chiral condensate and chiral
susceptibility of the p4fat7 action in the case of three degenerate flavors and $N_t=4$
are shown in Fig.~\ref{fig_pbp}. 
As expected the chiral transition becomes more pronounced for smaller quark
masses. The peak heights of the susceptibilities increase and a volume
dependence becomes visible.
We define the critical couplings ($\beta_c$) as the peak positions of the chiral
susceptibilities. However, we find that the peak positions of all
susceptibilities agree within our statistical accuracy. In
Fig.~\ref{fig_betac}(left) we show a comparison of $\beta_c$ for the two
actions and temporal lattice sizes analyzed so far.
The solid lines are fits to a power law. We find that the separation between the
$N_t=4$ and $N_t=6$ critical lines is much smaller for the p4fat7 action than
for the p4fat3 action. 
In fact, for small quark masses the critical couplings for $N_t=4$ and 6 
almost coincide which suggests the presence of a nearby bulk transition
in the parameter space defined by the p4fat7 couplings. This also leads to 
the onset of first order transitions at pion mass values larger than those 
found for the p4fat3 action. All this makes the p4fat7
action less suitable for studies of thermodynamics\footnote{We note that we
  have used a p4fat7 action with tree level coefficients and have not
  introduced tadpole improvement factors as it is done for the asqtad
  action \cite{Orginos:1999cr}.} as the extrapolation to the continuum limit
will become more difficult, i.e. will require larger $N_t$.

The two $(2+1)$-flavor lines shown in Fig.~\ref{fig_betac}(left) are
lines at constant strange quark mass. The lines corresponds to $am_s=0.065$ and
$am_s=0.1$ for the lower and upper curve respectively. A bare strange quark
mass of $am_s=0.065$ approximately corresponds to  the physical strange quark
mass.   

At each critical coupling $\beta_c$ we perform a zero temperature scale setting
calculation on a $16^3\times 32$ lattice, where we calculate hadron masses and
heavy quark potentials. From the potential we extract the Sommer scale $r_0$ as well
as the string tension $\sigma$ by fitting the lattice data with the usual Ansatz
$V(r)=-\alpha/r+\sigma r+const$. We show some preliminary results of the critical
temperature in units of the string tension in Fig.~\ref{fig_betac}(right). 
The quark mass dependence of the critical temperature for p4fat3, $N_t=4$ and
$n_f=3$ is consistent with the earlier calculations \cite{Karsch:2000kv}. 
The chiral extrapolation of $T_c/\sqrt(\sigma)$ with $(m_{PS}/m_{V})^2$
yields results consistent with earlier findings of
$T_c/\sqrt(\sigma)|_{m=0}=0.407(15)$ \cite{Karsch:2000kv}.
The chiral extrapolation for p4fat7 gives smaller values:
$T_c/\sqrt(\sigma)|_{m=0}\;\lsim\;0.4$ and $\lsim\;0.35$ for $N_t=4$ and $N_t=6$
respectively. Furthermore, the $N_t$ dependence
of the transition temperature is much stronger for p4fat7 
than for p4fat3. This again indicates a possible bulk transition in the
vicinity of the chiral limit of the p4fat7 action.

\section{Outlook}
The still ongoing determination of the critical temperature is part of
a detailed study of the thermodynamics of (2+1)-flavor QCD. As soon as we have
reliable chiral extrapolations for $N_t=4$ and $N_t=6$ we will add a $N_t$=8
calculation at selected quark mass values and try to perform the extrapolation
to the continuum limit. Moreover we will perform a study of the equation of
state with an almost realistic quark mass spectrum. 

\begin{theacknowledgments}
I thank all members of the RBC-Bielefeld Collaboration. This work has been
supported by the the U.S. Department of Energy under contract DE-AC02-98CH1-886.
\end{theacknowledgments}

\end{document}